\documentclass[twocolumn,useAMS]{mn2e_mine}
\usepackage{graphics,times}
\input{epsf}

\begin{document}

\newcommand{\beq}{\begin{equation}}
\newcommand{\eeq}{\end{equation}}

\def\dT{{\Theta'}}
\def\dU{{{\cal U}'}}
\def\dV{{{\cal V}'}}

\def\gamd{{\dot\gamma}}
\def\eps{{\epsilon}}
\def\cQ{{\cal Q}}
\def\cC{{\cal C}}
\def\cD{{\cal D}}
\def\cU{{\cal U}}
\def\cV{{\cal V}}
\def\cA{{\cal A}}
\def\cB{{\cal B}}
\def\bfn{{\bf n}}
\def\bfF{{\bf F}}
\def\ce{{\cal E}}
\def\ve{{\varepsilon}}
\def\cy{{\cal Y}}
\def\ck{{\cal K}}
\def\sgn{{\;{\rm sgn}}}
\def\ii{{\rm i}}
\def\dd{{\rm d}}

\def\sech{{\;{\rm sech}\;}}

\def\la{\hbox{\raise.35ex\rlap{$<$}\lower.6ex\hbox{$\sim$}\ }}
\def\ga{\hbox{\raise.35ex\rlap{$>$}\lower.6ex\hbox{$\sim$}\ }}
\def\beqa{\begin{eqnarray}}
\def\eeqa{\end{eqnarray}}
\def\sub#1{_{_{#1}}}
\def\order#1{{\cal O}\left({#1}\right)}
\newcommand{\sfrac}[2]{\small \mbox{$\frac{#1}{#2}$}}

\title{An idealized model for dust-gas interaction in a rotating channel}

\author[O.M. Umurhan ]{O.M. Umurhan\thanks{Email: mumurhan@physics.technion.ac.il},
\\
Department of Astronomy, City College of San Francisco, San Francisco, CA 94112, USA \\
Department of Geophysics and Planetary Sciences, Tel--Aviv
University, Israel \\
Department of Physics, Technion-Israel Institute of
Technology, 32000 Haifa, Israel}



\maketitle

\begin{abstract}
\small
A 2D model representing the dynamical interaction of dust and gas
in a planetary
channel is explored.
The two components are treated as interpenetrating fluids in
which the gas is treated as a Boussinesq fluid while the
dust is treated as pressureless.
The only coupling between both fluid states
is kinematic drag.
The channel gas experiences a temperature gradient
in the spanwise direction and it is adverse the constant force of
gravity.  The latter effects only the gas and not the dust component
which is considered to free float in the fluid.  The channel is
also considered on an f-plane so that the background vorticity
gradient can cause any emerging vortex structure to drift
like a Rossby wave.   A linear theory analysis is explored
and a nonlinear amplitude theory is developed
for disturbances of this arrangement. It is found that
the presence of the
dust can help generate and shape emerging convection patterns and dynamics
in the gas so long as the state of the gas exceeds a suitably defined
Rayleigh number appropriate for describing drag effects.
In the linear stage the dust particles collect quickly
onto sites in the gas where the vorticity is minimal, i.e. where
the disturbance vorticity is anticylonic which is consistent with previous studies.
The nonlinear theory shows that,
in turn,
the local enhancement of dust concentration in the gas
effects the vigor of the emerging convective roll
by modifying the effective local Rayleigh number of the fluid.  It is also
found that without the f-plane approximation built into the model
the dynamics there is an algebraic runaway caused by unrestrained growth in the dust concentration.
The background vorticity gradient forces
the convective roll to drift like a Rossby wave
and this causes the dust concentration
enhancements to not runaway.

\end{abstract}
\normalsize
\small
\section{Introduction and summary of results}
There has been growing interest in recent years in the possibility that
persistent vortex structures in protoplanetary discs may be the sites
where planetesimals are formed (Barge \& Sommeria, 1995,
Tanga {\it et al.}, 1996, Bracco {\it et al}, 1999, Barranco \& Marcus, 2000,
Barranco \& Marcus, 2006).
These investigators have put forth
the scenario in which a disc, laden with dust, supports some type
of long-lasting vortex (driven by some unspecified excitation mechanism,
except for Klahr \& Bodenheimer, 2003 and Barranco \& Marcus 2006)
that manages to attract the dust through the combined action of gas-particle
drag and Coriolis forcing (for example, Barranco \& Marcus, 2000).
Separately, it is also
well-known that strongly shearing flows (Keplerian discs qualify
under this classification) in rotating frames preferentially support
the persistence of anti-cyclonic structures while almost entirely
wiping out cyclonic vortices
\footnote{A coherent vortex
structure is said to be anti-cyclonic
if the signature of
the background vortex state in which it is embedded is
opposite the sign of the vortex structure itself.}
(Bracco, {\it et al.}, 1999).
In light of this, one of the
interesting results of the aforementioned dust-vortex interaction studies
is that anti-cyclonic vortices also happen to be
the sites onto which disc dust collects most rapidly (most
notably, Bracco {\it et al.}, 1999).
\par
In the investigations mentioned, the
dynamics of these dust-disc scenarios are treated as one-way:
the dust passively responds to the gas flow without any back-reaction
of the dust upon the gas.  Given that the current hypothesis of
protoplanetary discs is that their dust contributes no more than
a few percent of the total mass density of the disc (Hayashi, 1983),
it is reasonable to treat the dust as a collection of Lagrangian
tracers which responds to the vortex induced gas flow and having no
dynamical influence on the gas itself.
\par
However, it is interesting and instructive to turn the question
around and examine what would result if the dust does dynamically
effect the gas.  One could envision a situation in which either
the gas in the disc has been removed sufficiently or one is
investigating the dynamics taking place near the disc midplane,
where planetesimals are likely to be strongly concentrated.\par As
a thought experiment, suppose the gas in a model shearing
environment develops, through some type of dynamical instability,
a series of long-lived vortex structures. In addition to obviously
effecting the dust trajectory in the way it is usually envisioned,
it seems reasonable to suppose that the presence of dust would
dynamically influence, because of their mutual dynamic coupling
({\it e.g.} kinematic drag), the manner in which emerging gas
structures evolve.  This all would be plausible under conditions
in which the gas and dust have equal dynamical influence upon each
other.  Youdin and Goodman (2005) have conducted a similar study
considering the linear instability emerging from interpenetrating
streams of gas and dust whose only interaction is through kinematic
drag \footnote{There have been a number of studies of this sort
in which the interpenetrating streams are coupled to each other through
gravity.  For a more recent study see Bertin \& Cava (2006) and
references therein.}.
\par
Presented here is a simple and idealized model for the
way this type of interaction develops in a model confined environment.
The physical model employed here
is a flow in a rotating channel.  The material is treated
as two interpenetrating fluids of uniform density, one representing a pressureless dust ``fluid''
and the other an incompressible Boussinesq gas.  The gas thermodynamics is governed
by thermal conduction and we suppose that there is a constant gradient
of the gas temperature from one wall to the other (in the spanwise or
``wall-normal'' direction).  As part of the idealization that goes
into the model considered here,
the gas component \emph{alone} is subject to a constant force in
the wall-normal direction and that the Boussinesq buoyancy effects
are associated with this force in the usual way.
We suppose, further, that the gas
and dust are both subject
to an external force which gives rise to a linear Couette
shear of both fluids
in the direction parallel to the channel walls.
The coupling between the gas and dust is through a simple kinematic (Darcy) drag
prescription like, for example, in Tanga {\it et al.} (1995).
There is no viscosity.
The ingredients of this hypothetical
scenario already predisposes the gas
component susceptible to buoyant instability ala Rayleigh Benard.
Finally, because the dust fluid is treated as being
pressureless, no a priori restrictions upon the compressibility of the dust
is made.\par
In Section 2 we show the development of Rayleigh-Darcy type of convection.
In the formulation of the problem, the
dust density decouples from the linear analysis of the developing
convection roll.  A simple relationship is demonstrated that further shows
that a steady vortex
roll at the corresponding critical wavenumber causes the dust
density to grow algebraically when the roll is anti-cyclonic
and, conversely, the density is depleted where the roll is
cyclonic.\par
In Section 3 a nonlinear asymptotic analysis is developed
in the limit of large aspect ratio and under conditions of fixed thermal
flux on the channel walls and where the background vorticity is modeled
as an f-plane with a weak gradient.  In addition to this, the
imposition of a weak external force
will promote a steady flow exhibiting a weak amount of
shear (here it will be Couette).
The asymptotic analysis proceeds under the assumption that
both (i)
that the thermal time is much greater than the dust stopping time and, (ii)
that the time scale derived from the geometric mean of the thermal and
rotation times is also much greater than the stopping time of the dust.
The two conditions translates to a situation in which the dust stopping
time is actually much longer than the local rotation time.
This scaling regime also implies that the dust velocity
behaves as if it were irrotational at leading order.
\par
The model
shows algebraic instability (in the dust component) unless some amount of background
vorticity gradient is present
in the flow (as provided by the weak f-plane prescription).
This gradient
stabilizes the growth by promoting a Rossby wave drift of the
the convective roll (vortex) pattern.  Because the Rossby wave drift speed of the vortex
pattern is different than the background velocity field the effect here is for
the fluid pattern to drift past the dust component.  As given regions of the dust are exposed to
overall reductions of the total fluid vorticity (as measured with respect to the
background vorticity arising from the channel's rotation) the dust begins to collect
as expected from the behavior predicted by the linear theory.  However
because the pattern is not stationary the dust cannot locally collect ad infinitum because
the low vorticity part of the vortex pattern
will pass on through a given local dust region.  In turn, the dust region will
be exposed to patches of fluid passing by with enhanced vorticity and, consequently, will
experience a
reduction in the local dust concentration.  It is found, however, that secondary
growing
oscillations also appear in this formulation and these are wiped out when
a certain amount of diffusion in the dust concentration is added to the model.
It is also found that according to the model, places where the dust concentration is enhanced (depleted)
corresponds to zones where the convective roll is slightly weakened (strengthened).  We
show that this can be explained by an effective modification of the local Rayleigh number
due to the nonlinear rearrangement of the dust concentration.

\section{Two dimensional flow Equations}
Consider a two fluid description of a dust laden simple Boussinesq fluid
in a rotating channel.  For the sake of generality we suppose \emph{that only
the gas} phase has certain properties which cause it to be
dynamically influenced
by an external constant force of magnitude $g$ and with
direction normal to the channel walls ({\it viz.}
in ${\bf\hat y}$ direction).
In this toy model we additionally posit there to be a
second force represented by the term ${\cal F}_k$ which also
acts in the ${\bf \hat y}$ direction and its dependence
is on $y$ only: in other words,
${\cal F}_k = {\cal F}_k(y)$.
Unlike the other force $g$,
the force ${\cal F}_k$ affects both gas and dust.
The fluid is dynamically coupled to the dust because of
simple kinematic drag measured by the difference in
velocities between the gas and dust phase.
Finally, we assume that the thermodynamic transfer properties of the
gas are modeled by simple thermal conduction.  Thus, the equations of
motion for the gas phase are summarized to be,
\beqa
\nabla\cdot u &=& 0 \label{incompressible}\\
\rho\left(\frac{\partial}{\partial t} + u\cdot \nabla \right)u &=&
-\nabla P - \rho g {\bf {\hat y}} - \rho_d(u - u_d)/t_s \nonumber
\\
& &  + \rho\left\{ \begin{array}{c}f u^{(y)} {\bf {\hat x}}
\\{\cal F}_k - f u^{(x)} {\bf {\hat y}}\end{array}
\right \}
\label{gas:momentum}
\\
\rho C_p\left(\frac{\partial}{\partial t} + u\cdot \nabla
\right)T &=& K\nabla^2 T .
\eeqa
In which $u$ is the gas velocity and $u_d$ is the
dust velocity, $P$ is the gas pressure,
The dust density is $\rho_d$ while the gas density is simply $\rho$.
The superscript $(x)$ and $(y)$ appearing in front of velocity expressions
are respectively their $\hat x$ and $\hat y$ components.
The kinematic drag between phases is mitigated by the ``stopping-time''
$t_s$ and we assume it to be constant
 - see similar treatments by Cuzzi (1993) and Tanga {\it et al}, (1995).
If the channel rotates with rate $\Omega$ in direction ${\bf \hat z}$
then the Coriolis
term $f$ is given by $2\Omega$.
Because the gas phase is assumed to be Boussinesq its
thermodynamics is modeled by the usual expression of
a fluid undergoing fluctuations under constant pressure conditions
where $C_p$ is the specific heat at constant pressure.
\par

\par

Since this is a two-dimensional model, it will be more convenient for us
to consider the gas phase momentum equations in terms of
a single equation for the gas vorticity.
Since the gas phase is divergence free except when coupled to $g$,
we may consider the gas velocity as resulting from a streamfunction $\psi$
such that
\beq
\left[ u^{(x)},u^{(y)}\right] = \left[-\frac{\partial \psi}{\partial y},
\frac{\partial \psi}{\partial x}\right].
\eeq
It follows that by defining the vorticity as the curl of the fluid velocity,
$\omega = \nabla\times u$, then we have
\beq
\omega \equiv \frac{\partial u^{(y)}}{\partial x} -
\frac{\partial u^{(x)}}{\partial y} =
\frac{\partial^2 \psi}{\partial x^2} +
\frac{\partial^2 \psi}{\partial y^2}.
\label{vorticity:psi:linear}
\eeq
By taking the curl of (\ref{gas:momentum}) we find the following
\beqa
\rho\left[\frac{\partial \omega}{\partial t}
+ J(\omega,\psi)\right] &=&
-g\frac{\partial \rho}{\partial x}
-  \frac{1}{t_s}
\Bigl[\rho_d(\omega - \omega_d) \nonumber \\
&+& (u^{(x)} - u^{(x)}_d)\frac{\partial \rho_d}{\partial y}
- (u^{(y)} - u^{(y)}_d)\frac{\partial \rho_d}{\partial x}
\Bigr], \ \ \
\label{vorticity:full}
\eeqa
thereby eliminating direct reference to the gas phase's
divergence free nature and reducing the number of equations
from three to two.
The Jacobian appearing in (\ref{vorticity:full}) is defined
as
\[
J(f,g) \equiv \frac{\partial f}{\partial y} \frac{\partial g}{\partial x}
-\frac{\partial f}{\partial x} \frac{\partial g}{\partial y} .
\]
Unlike its gas counterpart, the dust phase is assumed to be
compressible, and this is why we have retained
all spatial dependences of $\rho_d$ in (\ref{vorticity:full}).
Furthermore, the dust phase vorticity, $\omega_d$, is defined from
the dust phase velocity, $u_d$, in the usual way, {\it viz.},
\beq
\omega_d \equiv \frac{\partial u^{(y)}_d}{\partial x} -
\frac{\partial u^{(x)}_d}{\partial y}
\eeq
\par
The dust phase is modeled as a pressureless, compressible fluid under the
influence of particle drag, the force ${\cal F}_k$ and the
Coriolis effect,
\beqa
\frac{\partial \rho_d}{\partial t}
+ \nabla\cdot(\rho_d u_d) &=& 0 \label{dust:cont} \\
\rho_d\left(\frac{\partial}{\partial t} + u_d\cdot \nabla\right)u_d &=&
\rho_d\left\{ \begin{array}{c}f u_d^{(y)} {\bf {\hat x}}
\\ {\cal F}_k - f u_d^{(x)} {\bf {\hat y}}\end{array}\right \} -
\rho_d(u_d - u)/t_s.
\eeqa
Since the underlying mass density states are presumed to be constant in both
fluids, it will prove beneficial to write departures of the dust density in terms of a fractional density parameter, $\delta$, defined by,
\beq
\rho_d = \bar\rho_d(1+\delta).
\eeq
\subsection{Steady State}
A thermal steady state is when the gas
conducts in the ${\bf \hat y}$ direction.  This means that there
is linear temperature gradient in $y$ given by,
$\bar T = T_* - \beta_T y$, in which $T_*$ is the background temperature
scale and where the constant $\beta_T$ represents,
in a sense, the steady background heat flux of the environment.
Denoting $\bar u$ and $\bar u_d$ as the steady
velocity profiles, a solution is $\bar u = \bar u_d$,
along with,
\beq f\bar u^{(x)} = f\bar u_d^{(x)} = {\cal F}_k(y),
\qquad \bar u_d^{(y)} =\bar u^{(y)} = 0, \quad {\rm and} \quad
\frac{\partial \bar P}{\partial y} = -\bar\rho g.
\label{steady_config}
\eeq
And as previously mentioned, the densities of both fluid states, $\bar \rho$
and $\bar \rho_d$, is assumed to be constant.
We consider steady conditions to have no ${\bf \hat x}$ direction dependences.
Thus, the temperature steady state condition is
\beq
K\frac{\partial ^2\bar T}{\partial y^2} = 0,
\eeq
which yields a steady flux state of the gas with a temperature profile
$\bar T = T_* - \beta y$, where $\beta$ is a constant related to the background
flux state of the gas and where $T_*$ is the background temperature scale.

\subsection{Linear Theory}
For the sake transparency, we consider linear disturbances of
the steady arrangement prescribed above when the force ${\cal F}_k$ is zero.
This means that
perturbations ensue under static conditions: $\bar u = \bar u_d = 0$.
Temperature fluctuations are represented by
\beq
T = T_* - \beta y + \theta,
\eeq
where $\theta$ represents the temperature disturbance away from
the static state.
Except for buoyancy effects, the gas density is constant. Therefore
the Bousinessq approximation says that,
\beq
\rho' = \rho - \bar\rho = -\alpha \theta,
\eeq
where here the coefficient of thermal expansion at constant pressure is
represented by $\alpha$.
A prime on any density quantity is
meant to designate its deviation from the steady uniform state,
which will be designated with an overbar over the quantity.
\par
We suppose that all dynamical disturbances occur on a length
scale of $d$.  Consequently spatial scales are nondimensionalized
on $d$ and, furthermore, they are written as
\beqa
x = \xi d &,& \qquad y =  Y d \nonumber \\
\frac{\partial}{\partial x}
\rightarrow  \frac{1}{d}\frac{\partial}{\partial \xi }
&,&
\frac{\partial}{\partial y}
\rightarrow  \frac{1}{d}\frac{\partial}{\partial Y} \nonumber
\eeqa
$\tau = C_p \rho d^2/K$ is a thermal time scale and, as such, we
scale all temporal derivatives by it.
Temperature disturbances are scaled by $\beta d$ and is written in
nondimensional terms as $\theta = \Theta \beta d $.
Velocities scale with $d/\tau$ which
sets the scaling for both the stream function and vorticity (for both
fluid phases),
\[
\psi = \frac{d^2}{\tau} \tilde\psi, \quad
\omega = \frac{1}{\tau}\tilde\omega,
\]
in which $\tilde\psi$ and $\tilde\omega$ are the nondimensional stream function
and vorticity, respectively.
In this formalism the nondimensional linearized perturbation
equations for the gas are,
\beqa
\left(\frac{\rho}{\rho_d}\right)
\frac{1}{Pr}
\frac{\partial\tilde\omega}{\partial t}
&=& R\frac{\partial\Theta}{\partial\xi} -
(\tilde\omega-\tilde\omega_d)
\label{gas:vorticity:linear}
\\
\frac{\partial\Theta}{\partial t}
&=& \frac{\partial\tilde\psi}{\partial\xi} +
\left(\frac{\partial^2}{\partial \xi^2} +
\frac{\partial^2}{\partial Y^2}
\right)\Theta
\label{gas:thermal:linear} \\
\tilde \omega &=& \left(\frac{\partial^2}{\partial \xi^2} +
\frac{\partial^2}{\partial Y^2}
\right)\tilde\psi.
\label{vorticity:psi:linear}
\eeqa
Whilst for the dust phase,
\beqa
\frac{\partial}{\partial t}
\left(\frac{\rho'_d}{\bar\rho_d}\right)
+ \tilde C &=& 0 ,
\label{dust:continuity:linear}
\\
\frac{1}{Pr}
\frac{\partial\omega_d}{\partial t} &=& - (\omega_d-\omega)
- \tilde f \tilde C ,
\label{dust:vorticity:linear}
\\
\frac{1}{Pr}
\frac{\partial \tilde C}{\partial t} &=& - \tilde C + \tilde f \omega_d.
\label{dust:dilitation:linear}
\eeqa
The variable $\tilde C$, defined by
\beq
\tilde C \equiv \frac{\partial \tilde u_d^{(x)} }{\partial\xi}
+  \frac{\partial \tilde u_d^{(y)}}{\partial Y},
\label{tilde_C_def}
\eeq
represents the dilatation of the dust phase.  The term
$\tilde u_d$ represents the dust velocity scaled to $d/\tau$.
The Rayleigh number, $R$, is defined to be
\beq
R = \frac{1}{\bar\rho}g\alpha\beta_T t_s
\left(\frac{C_p\bar\rho d^2}{K}\right)\left(\frac{\bar\rho}{\bar\rho_d}\right),
\label{def:Rayleigh}
\eeq
while an effective Prandtl number $Pr$ is identified with
\beq
Pr = \frac{\tau}{t_s} = \frac
{C_p\bar\rho d^2}{Kt_s}.
\eeq
Finally, the Coriolis parameter is
scaled to the stopping time: $\tilde f \equiv ft_s$.\par
We note that in the dust momentum equation, out of which both
(\ref{dust:vorticity:linear}) and
(\ref{dust:dilitation:linear}) are formed,
all appearances of the dust density and its fluctuation, $\rho'_d/\bar\rho_d$,
explicitly drop out,
In practice this means that
(\ref{dust:continuity:linear}) explicitly decouples in
a normal mode
analysis of the perturbation equation set
(\ref{gas:vorticity:linear})-(\ref{dust:dilitation:linear}).
This observation has an interesting consequence for the nonlinear theory
in the upcoming section.
\par
The limited coupled set of equations
(\ref{gas:vorticity:linear})-(\ref{vorticity:psi:linear}) and
(\ref{dust:vorticity:linear})-(\ref{dust:dilitation:linear}) admit
steady ({\it i.e.} $\partial/\partial t = 0$) solutions.
The particle phase variables are simply related to the gas
phase variables by,
\beqa
\tilde\omega_d &=& \frac{\tilde \omega}{1 + \tilde f^2}
\label{eq:lin:1} \\
\tilde C &=& \frac{\tilde f \tilde \omega}{1 + \tilde f^2},
\label{eq:lin:2}
\eeqa
Note that (\ref{eq:lin:2}) says that the dust dilatation is
related to the gas vorticity, which will have consequence shortly.
The gas phase equations become,
\beqa
0 &=& R\frac{\partial \Theta}{\partial\xi} - (\omega-\omega_d),
\label{eq:lin:3}
\\
0 &=& \frac{\partial \tilde\psi}{\partial\xi}
+\left(\frac{\partial^2}{\partial\xi^2} +
\frac{\partial^2}{\partial Y^2}\right)\Theta,
\label{eq:lin:4}
\eeqa
along with (\ref{vorticity:psi:linear}) unaltered as it appears.
Combining (\ref{vorticity:psi:linear}) with (\ref{eq:lin:1}-\ref{eq:lin:4})
yields a single equation for $\tilde\psi$,
\beq
\left(\frac{\partial^2}{\partial Y^2} +
\frac{\partial}{\partial \xi^2}\right)^2\tilde\psi +
R\frac{1+\tilde f^2}{\tilde f^2}
\frac{\partial^2  \tilde\psi}{\partial \xi^2} = 0
\label{linear:equation:psi}
\eeq
(\ref{linear:equation:psi}) is supplemented with
boundary conditions.
As for the thermal conditions: we explore two different types of
requirements, namely that the (i) the temperatures are fixed at the
boundaries or (ii) the thermal fluxes are fixed at the boundaries.
The former condition is the more familiar of the two in which, in practice,
it requires that the temperature fluctuations are zero at the boundaries.
The second condition is, effectively, thermal insulation and it, in practice,
amounts to setting $\partial\Theta/\partial Y = 0$ at the boundaries.
\par
Assuming, in general for all quantities,
periodic solutions in the horizontal $\xi$
direction with corresponding horizontal wavenumber $q$, {\it e.g.}
\[
\tilde\psi = \bar\psi(Y)\exp{iq\xi} + {\rm c.c.},
\]
we find that
for fixed-temperature boundary conditions
the lowest normal mode solution in $Y$ for $\tilde\psi$ is
\beq
\tilde\psi  \sim \exp {i q \xi} \cos \pi Y ,
\eeq
where the system eigenvalue, which is an effective Rayleigh number defined
by,
\beq
R^{{\rm eff}} \equiv \left(\frac{1 + \tilde f^2}{\tilde f^2}\right) R,
\eeq
must satisfy,
\beq
R^{{\rm eff}} =
\frac{(\pi^2 + q^2)^2}{q^2}.
\label{effective:Rayleigh:number:FT}
\eeq
There is a minimum value of $R^{eff}$
at a critical wavenumber $q_c$
required for such marginal
solutions to exist. We denote the minimum value of the effective
Rayleigh number as
$R^{eff}_c$ and we find that for fixed-temperature boundary
conditions $q_c = \pi$ and,
$R^{eff}_c = 4\pi^2 \approx 40$ - which is the number expected
for Rayleigh-Darcy convection (Horton \& Rogers, 1945).
\par
\begin{figure}
\begin{center}
\leavevmode
\epsfysize=3.6cm
\epsfbox{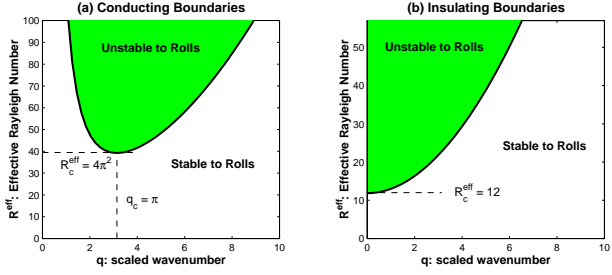}
\end{center}
\caption{{\small Critical Rayleigh number as a function of
scaled horizontal wavelength $q$. (a) Conducting boundaries and (b)insulating
(fixed-flux) boundaries. }}
\label{fig:critical_rayleigh}
\end{figure}

The solution for perfectly insulating boundaries resists the straightforward
form characterizing the fixed temperature result.  It proves to be easier
to seek numerical solutions here instead.  Figure \ref{fig:critical_rayleigh}(b)
 displays
the Rayleigh number as a function of horizontal wavenumber for
these marginal conditions.  It turns out that
$R^{{\rm eff}}_c = 12$ and it occurs at $q=0$.  The result that
the critical Rayleigh number occurs at infinite horizontal
scales, though curious, is a well known result in standard Rayleigh-Benard
problems and its implications
have been studied by many investigators (e.g. Chapman \& Proctor, 1980).
\par
Irrespective of the thermal boundary condition assumed the marginal
roll state can cause the dust density to grow algebraically.  To see this
consider (\ref{dust:dilitation:linear}) with the dust dilatation term
re-expressed in terms of the relationship
(\ref{eq:lin:2}).  It is found that the time rate of change of
the dust density is expressed by
\beq
\frac{\partial}{\partial t} \left(\frac{\rho_d'}{\bar\rho_d}\right)
= -\frac{\tilde f \tilde \omega}
{1 + \tilde f^2}.
\label{dust:growth}
\eeq

In other words, the dust density changes linearly in time (since
there is no time dependence in $\tilde\omega$) and, additionally,
this change is fastest at places where the vortex amplitude is extremal.
Given the relationship
in (\ref{dust:growth}) it is apparent that dust density grows
fastest near anti-cyclonic vortex perturbations
: this is clear since the RHS of
the expression is positive only if the product $-\tilde f \tilde\omega$
is positive and this can only happen if the signs of $\tilde f$ and
$\tilde\omega$ are opposite.
\par
To
be more accurate in this description of growing dust concentration it should
be stated that
if the sign of the growing vortex
perturbation
is negative it only means that this part of the fluid state starts
appearing to have
lower vorticity than the original global state.
The result of this linear theory, taken in this light,
is not too surprising because earlier work on the behavior
of Lagrangian tracer particles in fully developed isotropic turbulence flow
(Squires and Eaton, 1991) show that particles concentrate in those parts
of the flow which are regions of lower vorticity and high strain rate.

\begin{figure}
\begin{center}
\leavevmode
\epsfysize=7.0cm
\epsfbox{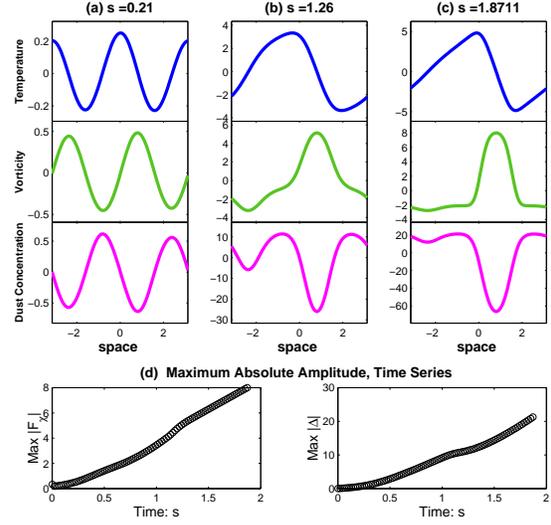}
\end{center}
\caption{{\small Representative profiles for Model A:
$\tilde\beta = \tilde b_1 = 0$ and $r_2 = 5.30$ and $\tilde \phi = 10$.
Plotted are the lowest order temperature amplitude $F$,
the lowest order vorticity amplitude $F_\chi$,
followed by the dust concentration, $\Delta$.
Panel (a), at time $s\sim 0.21$.
Panel (b), time $s\sim 1.25$.
Panel (c), blow up time $s\sim 1.87$.
The scaled dust concentration always seems to trace the gas
vorticity $F_\chi$ even throughout the blow up time.
Panel (d) is the time series for the absolute values of vorticity
scale and dust concentration, respectively.}}
\label{fig:barebones}
\end{figure}

\section{An Asymptotic Nonlinear Model}
In order to understand how to handle the curious situation
involving the algebraic growth of the particle density perturbation
even though the gas phase is neither growing nor decaying
we asymptotically analyze the governing equations
under certain extreme conditions.
In particular, we will explore the nonlinear development of
disturbances for fixed thermal flux boundary conditions.  Furthermore
we will assume that the Prandtl number is much larger than the
scaled Coriolis parameter, $\tilde f$ which will also be much greater
than one, {\it viz.},
\[
Pr \gg \tilde f \gg 1.
\]
This extreme ordering says physically that
the dust stopping time, $t_s$, is much longer than the local rotation
time $\sim \Omega^{-1}$.  This scaling however also says
that the dust stopping time is much less than the
geometric mean of the thermal and rotation times.
In other words,
\[
\frac{\tau}{\Omega} \gg t_s^2,
\qquad{\rm and} \qquad
\Omega t_s \gg 1.
\]
For the sake of this asymptotic calculation we take $Pr \rightarrow \infty$.
\par
Furthermore we will suppose that the
horizontal lengths scales are much larger than the scales in
direction of gravity and we will introduce the small scaling parameter
$\epsilon$ to represent this extreme state.  Corresponding to this
situation we introduce the scaled horizontal variable $X = \epsilon \xi$,
where $X$ is an order one quantity.  Additionally we introduce
an order 1 scaled time variable $T$ such that $T = \epsilon^4 t$.
Thus, we replace all derivative operators with
\beq
\frac{\partial}{\partial \xi} \rightarrow \epsilon\partial_X,
\quad
\frac{\partial}{\partial t} \rightarrow \epsilon^4\partial_T
\eeq
From now on we will appeal to the shorthand notation for
derivative operations for compactness of notation
and clarity of derivation.  Correspondingly, we find that
a distinguished limit exists when the stream function is
$\order\epsilon$ the temperature fluctuation.  Specifically, we
say that
\[ \tilde\psi = \epsilon\Psi,\qquad
\tilde\omega = \epsilon\Omega,
\]
where $\Psi$ and $\Omega$ are now order 1 quantities.\par
We introduce the external force, ${\cal F}_k$, into the analysis
in which its nondimensionalization
is expressed through the relationship,
\[
{\cal F}_k(y) = \frac{d}{\tau^2}\tilde{\cal F}_k(Y),
\]
where $\tilde{\cal F}_k$ is nondimensional.  We assume that the steady
external force has the form
\beq
\tilde {\cal F}_k = \frac{1}{\epsilon^4}b\phi Y,
\eeq
where $b$ and $\phi$ are order 1 constants.
\par
In order to achieve proper asymptotic balance
we exploit the freedom we have in choosing the ``largeness'' of
the Coriolis
parameter by scaling it (ad hoc) as
\beq
\tilde f = \frac{1}{\epsilon^5}\phi
\left(1 + \frac{\beta}{\phi}\epsilon^6 Y\right)
\eeq
where $\phi$ is, as before, order 1.
The bizarre choice behind this asymptotic ordering
is motivated by introducing the dust disturbance
term $\tilde\delta$ into the subsequent analysis at the right stage.
The $\beta Y$ term, which tacitly represents an assumption of
a weak f-plane effect (see Pedlosky), is introduced a priori in order
to ensure stability of the nonlinear model.
Though it is an artificial device, it has the effect of causing
a Rossby wave type of drift of a vortex pattern.
\par
With respect to (\ref{steady_config}), the resulting
steady x-direction flow is weakly Couette.  Denoting
the steady flow by $U$ and $U_d$ then we have
\beq
U = U_d = \epsilon b Y + \order{\epsilon^7}.
\label{steady_shear}
\eeq
From here on out, we consider disturbances of the dust and gas
velocities on top and over this base state.
\par
With these scalings in place we find that the dust quantities
require the following scalings in order for there to be nontrivial
balance:  the fractional dust density scales as $\epsilon^2$ and
from here on out it is written as
\beq
\frac{\rho'_d}{\bar\rho_d}
= \epsilon^2\delta.
\eeq
We find that the dust dilatation is first nontrivial at order $\epsilon^6$
and it is written as,
\beq
\tilde C = \epsilon^6 C_6 + \order{\epsilon^8}
\label{C6_order}
\eeq
whereas the dust vorticity scales even smaller
such that it is nontrivial at order $\epsilon^{11}$! In other words,
\beq
\tilde\omega_d = \epsilon^{11}\Omega_d.
\eeq
Because of this last observation we will, for all intents
and purposes, treat the dust velocity as derived from
a potential which similarly scales
on order $\epsilon^6$.  In other words, $u_d = \epsilon^6\nabla \Gamma$.
As such, each component
of the dust velocity scales as
\beq
\tilde u_d^{(y)} = \epsilon^6 v_6
 = \epsilon^6 \frac{\partial \Gamma}{\partial Y},
\qquad
\tilde u_d^{(x)} = \epsilon^7 w_7
= \epsilon^7 \frac{\partial \Gamma}{\partial X}.
\eeq
Together with the scaling relationship (\ref{C6_order}) and general definition
(\ref{tilde_C_def}) it is clear that at these initial
orders the dust velocity satisfies a Poisson type of equation,
\beq
\nabla^2\Gamma = C_6.
\eeq
\par
With these scalings and assumptions
in place we consider the temporal development of disturbances when
the Rayleigh number deviates from its critical value by an order $\epsilon^2$
amount written as,
\beq
R \rightarrow R_0 + \epsilon^2 R_2.
\eeq
For this analysis we will consider the development in terms of fixed-flux
boundary conditions which means that the critical Rayleigh number about
which we will be expanding around will be 12.  However, the following
calculation self-consistently selects the proper value of $R_0$.  The
governing equations (in the $Pr \rightarrow \infty$ limit)
with these scalings inserted now reads for the gas phase,
\beqa
R_0\left(1 + \epsilon^2\frac{R_2}{R_0}\right)\partial_X\Theta
- (1+\epsilon^2\delta_2)\Omega & = & \epsilon^2\beta\partial_X\Psi
\label{ap:mom:body}
\\
\epsilon^4\partial_T\Theta + \epsilon^2{\cal J}(\Theta,\Psi)
+ \epsilon^2 b_1 Y \partial_X\Theta & = &
\epsilon^2\partial_X\Psi + \epsilon^2\partial_X^2\Theta + D^2\Theta
\label{ap:heat:body}
\\
\Omega &=& \left(D^2 + \epsilon^2\partial_X^2\right)\Psi,
\label{ap:stream:body}
\eeqa
and for the particle phase they are,
\beqa
0 &=& \epsilon\Omega -\epsilon\phi C_6 + \order{\epsilon^7}
\label{ap:particle_divergence:body}
\\
0 &=& -\epsilon^6 C_6 + \epsilon^6 \phi\Omega_d + \order{\epsilon^{12}} \\
\epsilon^6\partial_T\delta + \epsilon^6C_6
& = &
0 + \order{\epsilon^8}
\label{ap:particle:body}
\eeqa
The new Jacobian symbol, ${\cal J}$ is defined on the variables
$X$ and $Y$.  In the next section we set out to solve, order by order,
the set of equations (\ref{ap:mom:body}-\ref{ap:particle:body}) with fixed flux
thermal boundary conditions.
\subsection{Expansions}
The set of equations (\ref{ap:mom:body}-\ref{ap:particle:body})
are solved as a perturbation series expansion in powers of $\epsilon^2$.
Thus, the following expansions are presumed for this purpose,
\beqa
\Theta = \Theta_0 + \epsilon^2\Theta_2 + \epsilon^4\Theta_4 + \cdots \\
\Psi = \Psi_0 + \epsilon^2\Psi_2 + \cdots \\
\Omega = \Omega_0 + \epsilon^2\Omega_2 + \cdots \\
\delta = \delta_2 +  \epsilon^2\delta_4 +  \cdots
\eeqa
(\ref{ap:heat:body}) to lowest order becomes simply,
\beq
D^2 \Theta_0 = 0.
\eeq
The solution of this equation subject to the condition that the
flux remain fixed on the boundary is,
\beq
\Theta_0 = f(X,T),
\eeq
meaning to say that the lowest order temperature perturbation
is independent of the vertical coordinate.  The rest of the
expansion procedure is standard and we relegate its full
exposition to Appendix \ref{asymptotix}.  What results are
two coupled evolution equations for the temperature
disturbance $f$ and the lowest order particle density $\delta_2$
in (\ref{delta:evolution}) and (\ref{f:evolution}).  These
are reproduced below,
\beqa
f_T & = & - \kappa f_{XXXX} + \mu(f^3_X)_X
- \left[(r_2 - \delta_2)f_{X}\right]_X \nonumber \\
& & -b_1 S (f_X^2)_X - \beta B f_{XXX},
\label{f:eq}
 \\
\delta_{2T} & = &
-\phi R_0 f_X.
\label{delta:eq}
\eeqa
The positive constants $\kappa,\mu,B,S$ are defined in
Appendix \ref{asymptotix}.  $r_2$ is an effective
departure from the critical Rayleigh number $R_0=12$
given by
\[
r_2 = \frac{R_2}{12} - \frac{b_1^2}{120}.
\]
The parameter $r_2$ can be interpreted as representing
an effective departure from criticality which includes the
fact that a constant (linear) shear is a stabilizing influence against the onset
of rolls in the linear regime.

\begin{figure}
\begin{center}
\leavevmode
\epsfysize=7.0cm
\epsfbox{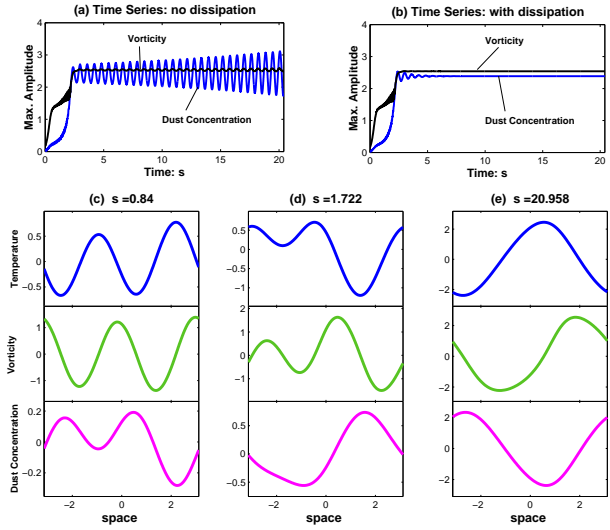}
\end{center}
\caption{{\small Representative profiles for Model B:
$\tilde b_1 = 0$ and $r_2 = 5.30$, $\tilde \phi = 10$,
$\tilde \beta = -10$.
(a), maximum amplitude of the dust concentration and
the scaled vorticity when there is dissipation (${\cal B} = 0$).
The amplitudes oscillate with growing size on a long-time scale.
(b), maximum amplitude of the dust concentration and
the scaled vorticity with dissipation present, (${\cal B} = 1$).
The oscillations are quelled while the overall average maximum amplitude
of the structures are preserved.
Panels (c-e) representing profiles of $F,F_\chi,\Delta$
at various stages in the ensuing evolution.  It should be observed that
the dust concentration, $\Delta$,
eventually locks onto the temperature profile, $F$,
at late times.}}
\label{fig:no_shear_1}
\end{figure}

\subsection{Canonization and Identification of Effects}
It is more convenient to define a number of rescalings of
(\ref{f:eq}) and (\ref{delta:eq}) in order to transparently
display the structure of the two coupled equations.
One may define new space and time coordinates, $\xi$ and $s$
along with new amplitude scalings $F$ and $\Delta$ by,
\beqa
& & T \rightarrow s\kappa,\qquad X \rightarrow \sqrt\kappa \chi, \nonumber \\
& & f(X,T)
\rightarrow \left(\frac{\kappa}{\mu}\right)^{1/2} F(\chi,s),
\qquad \delta_2(X,T) \rightarrow \Delta(\chi,s).
\label{canonization}
\eeqa
Consequently, (\ref{f:eq}) and (\ref{delta:eq}) rewritten
in terms of the definitions (\ref{canonization})
is
\beqa
F_s & = & - F_{\chi\chi\chi\chi} + (F_\chi^3)_\chi
-\Bigl((r_2-\Delta)F_\chi\Bigr)_\chi \nonumber \\
& & \ \ \ \
- \tilde b_1 (F_\chi^2)_\chi
- \tilde \beta  F_{\chi\chi\chi}, \label{theta:evol}\\
\Delta_s & = & -\tilde\phi F_\chi, \label{dust:evol},
\eeqa
where
\[
\tilde b_1 \equiv b_1\frac{S}{\mu}, \qquad
\tilde \beta \equiv \beta \frac{B}{\sqrt\kappa}, \qquad
\tilde\phi \equiv \phi R_0 \frac{\kappa}{\sqrt{\mu}}.
\]
The model set (\ref{theta:evol}) and (\ref{dust:evol}) are now
in a more transparent form and allows us to discuss the
effects they contain.  The role of the first two terms on the RHS of
(\ref{theta:evol}) are well known ({\it see} Chapman \& Proctor, 1980)
respectively, (i)
to damp the temperature profile due to horizontal thermal dissipation
and (ii) to control amplitude due to nonlinear advection of temperature.
The $r_2$ expression in the third term on the RHS of
(\ref{theta:evol}) constitutes the usual source of amplitude growth
because of positive departures from the critical Rayleigh number.  The
$\Delta$ term expresses the fact that as the particle concentration grows
the local Rayleigh number of the fluid goes down somewhat.  This is
best seen by observing the definition of the Rayleigh number
(\ref{def:Rayleigh}): if the dust density $\rho_d$ goes up then
$R$ goes down and vice versa.
The other terms on the RHS of
(\ref{theta:evol}) represent, respectively, (iii) the nonlinear twisting of
the temperature profile due to shear
and (iv) an effective Rossby wave type of drift of an emerging profile.
As discussed in the previous section, (\ref{dust:evol}) represents the
local enhancement of dust concentration
due to the emerging convection roll's vorticity, which is, to lowest
order, proportional
to $F_\chi$ as demonstrated in Appendix \ref{asymptotix}.
\par
Note that the nonlinear term $(F_\chi^2)_\chi$
associated with the shear is known
to promote subcritical transitions
in Rayleigh-Benard convection (Gertsberg \& Sivashinsky, 1981)
and will do so here in this model as well but we
will not explore this feature in this study.  This
term appears generically so long as there is some sort of
symmetry-breaking effect in the system (e.g. see Depassier \& Spiegel, 1982).

\begin{figure}
\begin{center}
\leavevmode
\epsfysize=3.cm
\epsfbox{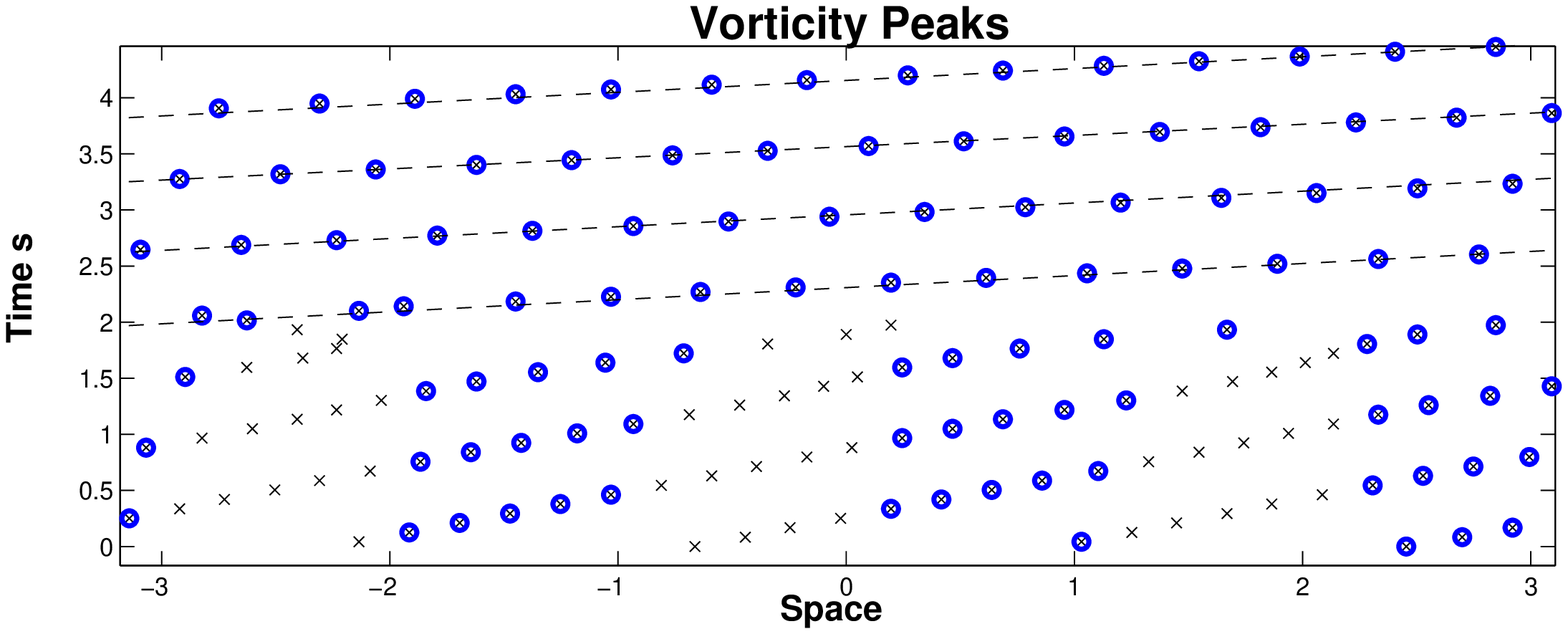}
\end{center}
\caption{{\small Position of the peaks of the vorticity profile for
Model B. The maximum peak is labelled with a circle and all secondary
peaks are denoted with x's.  After the transient readjustment
phase is over (near $s\sim 2$) the system goes from multiple peaks
to only one.  After this time, the path of the peak is traced out
with a dashed line and it is evident that it propagates from left
to right on this spacetime diagram.}}
\label{no_shear_vorty_peak}
\end{figure}

\begin{figure}
\begin{center}
\leavevmode
\epsfysize=2.75cm
\epsfbox{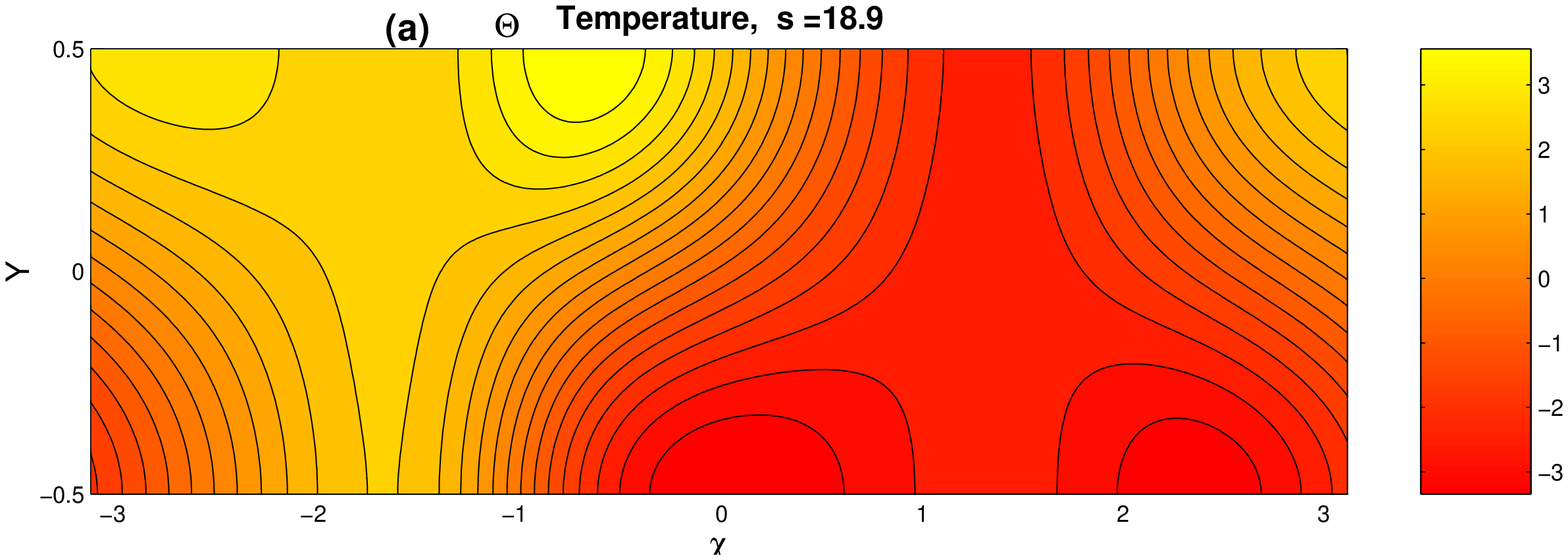}
\end{center}
\label{fig:no_shear_temperature}
\begin{center}
\leavevmode
\epsfysize=2.75cm
\epsfbox{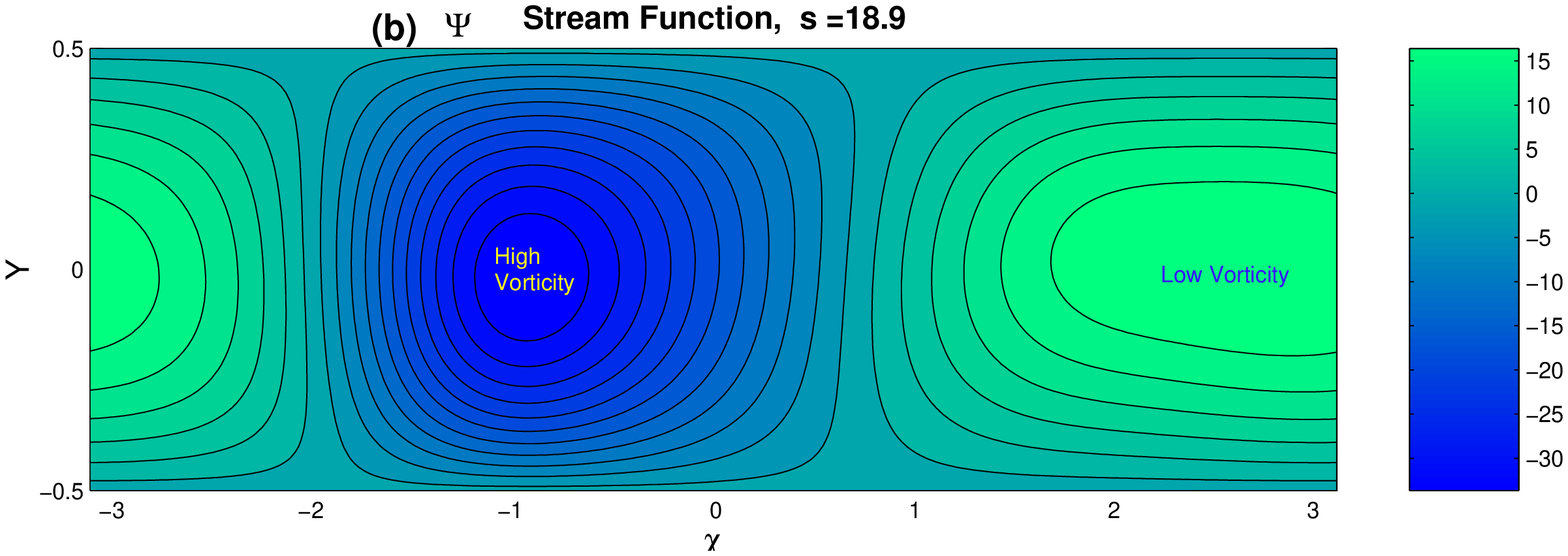}
\end{center}
\caption{{\small Representative contours for Model B when the
growth has settled onto the steady state.  For these contours $\epsilon = 1$.
(a) Temperature fluctuation contours, $\Theta$.
(b) Stream function contours, $\Psi$.}}
\label{fig:no_shear_stream}
\end{figure}

\subsection{Selected Solutions}
The reduced asymptotic equations (\ref{theta:evol}-\ref{dust:evol}) are evolved
according to the numerical spectral scheme outlined in Appendix
\ref{numerical_method} and is similar to the tactics used in Umurhan \& Regev (2004).  The
robustness of the scheme was checked against the analytic solutions of
a similar evolution equation derived by Chapman and Proctor (1980).
The
tests demonstrated agreement between the numerical scheme and the
analytical results to one part in $10^{-7}$.  The results reported here
were generally computed on 256 Fourier modes and when convergence was
doubted, the results were checked against simulations
based on 512 and 1024 modes.
Because there are
fourth order derivatives in the linear terms of (\ref{theta:evol})
we note that, (a) no artificial viscosity was needed and,
(b) small time steps on the order of $dt = 10^{-4}$ were typically taken.
All simulations started out with random white noise for $F$ whose total
amplitude did not exceed $10^{-2}$.  The dust concentration always started
out with zero initial amplitude.
\par
We consider here four selected models to discuss;
\begin{itemize}
\item \textbf{Model A.} No background vorticity
gradient and no shear, $\tilde b_2 = \tilde\beta = 0$, while $r_2 = 5.3$ and
the dust coupling is set $\tilde\phi = 10$,
\item \textbf{Model B.} The background vorticity
gradient is negative,$\tilde\beta = -10$, there is no shear
$\tilde b_2 =0$, along with $r_2 = 5.3$ and
$\tilde\phi = 10$ and,
\item \textbf{Model C}
Some amount of shear with a prograde sense, $\tilde b_1 = -10.0$
along with the same values of $r_2$,$\tilde\phi$ and $\tilde\beta$ of
Model B.
\end{itemize}
It should be noted that in all
models the background vorticity is getting smaller with increasing
$Y$.  Additionally, the results reported here required there to be
some dissipation added to the equation describing the dust concentration
evolution or otherwise the solutions blow up.  This is discussed in
more detail for Model B.

\subsubsection{Model A}
This and other models like it, where $\tilde\beta = 0$,
are unstable.  The amplitude and dust
concentration steadily runaway once the dust concentration locks into
the standing vorticity profile that emerges.
To see this more readily refer to Figure \ref{fig:barebones}.
At early times in the
simulation the local dust concentration grows fastest wherever
$G = F_\chi$, which is proportional to the leading order vorticity,
is greatest. This
behavior is in line with the conclusions of the linear theory section.
For $r_2 = 5.3$ and $\tilde\phi = 0$ ,
the fastest growing wavenumber is $n=2$ and this readily shown
in the linear theory analysis of (\ref{theta:evol}).
Despite the linear theory expectations however, the final roll profile will
go from being double humped to being single humped
as was shown by Chapman and Proctor (1980).  This is because the nonlinear
stability analysis shows that the system
prefers to eventually settle onto a profile with a single roll.
\par With
$\tilde\phi$ non-zero, the same sort of late time behavior is observed.
The dust concentration also qualitatively resembles the vorticity distribution
and it, similarly to the linear theory, appears to grow/deplete without bound.
Those places where the vorticity amplitude is greatest are those places
where the dust concentration is greatly depleted.  Consequently, those
locations in the gas correspond to greatly enhanced values of the local
Rayleigh number thereby causing the roll profile there to continue growing,
in what appears to be, without bound.
Naturally, the validity of this
predicted blowup should be questioned
since unbounded growth means the asymptotic solution
has ceased being valid.

\subsubsection{Model B: Vorticity Gradient with No Shear}
Introduction of $\tilde\beta$ stabilizes the runaway
growth seen in Model A.  The amplitude of the emerging
convection roll settles onto the vicinity of a given
value though on oscillation sets in.  On a very long time
scale this oscillation, which shows up in both the
roll amplitude and dust concentration amplitude,
exhibits a growth in amplitude and it eventually blows up
as is evidenced in Figure \ref{fig:no_shear_1}a.
This happens for all models in which $\tilde\phi$ is not
zero and the secular growth time scale is shorter as
$\tilde\phi$ becomes larger.  In order to stabilize the
long time behavior a dissipation term is introduced into the
dust evolution equation so that, and from here on out, instead
of (\ref{dust:evol}) the following equation is evolved,
\beq
\Delta_s  =  -\tilde\phi F_\chi
+{\cal B}\Delta_{\chi\chi},
\label{dust:evol:brownian}
\eeq
${\cal B}$ is the scale of the dissipation and in the simulations
it is taken to be unity.
With the simulation
run with ${\cal B} = 1$, the resulting maximum absolute amplitude is
shown in
Figure \ref{fig:no_shear_1}b.  The oscillations observed
for ${\cal B} = 0$ are removed and the overall average amplitude of the
underlying structures are preserved.
\par
The time evolution of these simulations begin with an initial
growth phase.  For $r_2 = 5.3$, the linear theory predicts
that the fastest growing Fourier mode is $k=2$ and, as such,
in the early stages the temperature profile is double humped
as in Figure \ref{fig:no_shear_1}b.
As Chapman and Proctor (1980) demonstrated,
this type of profile is nonlinearly unstable and
eventually the system undergoes a phase of
transient readjustment (around $s = 1.5$) in which the temperature
and vorticity go from being a two-humped profile down to a one-humped profile.
Figure \ref{fig:no_shear_1}c shows the state of this profile readjustment
near the time $s \sim 1.7$.
\par
Additionally, Figure \ref{no_shear_vorty_peak}
demonstrates how the profile drifts with increasing $\chi$ as
a function of time.
In this steady drifting state, the profiles are single
humped and, as in Figure \ref{fig:no_shear_1}d,
after the post readjustment phase
the dust concentration profile locks onto the temperture
profile tightly.
The reason that the dust concentration does not grow without
bound when the roll pattern drifts is simple.  In Model A, because
$\tilde\beta = 0$, the temperature and vorticity
profile that develops does so in place.
Because the dust concentration responds to local variations
of the vorticity,
places in the growing roll profile
where the vorticity is most reduced are also
where the rate of dust collection is greatest.
Without a suitable saturation mechanism
the greatly enhanced dust concentration begins to modify the
local Rayleigh number of the fluid.  This, in turn, causes the
roll amplitude to grow with ferocity.  With a larger amplitude in the
vorticity will come an ever increased concentration of
dust.  In this way the cycle feeds on itself and it runs away.
On the other hand, when the roll profile drifts, local enhancements of
dust do not grow without bound because
the low vorticity patch of the fluid that causes the dust to concentrate
will eventually drift past.
The local dust will then encounter gas with a
higher than normal degree of vorticity and this will cause the
the dust concentration to begin to diminish.
\par Finally, Figures
\ref{fig:no_shear_stream}(a-b) show the temperature and
stream function in the full 2D domain.
\par


\begin{figure}
\begin{center}
\leavevmode
\epsfysize=7.0cm
\epsfbox{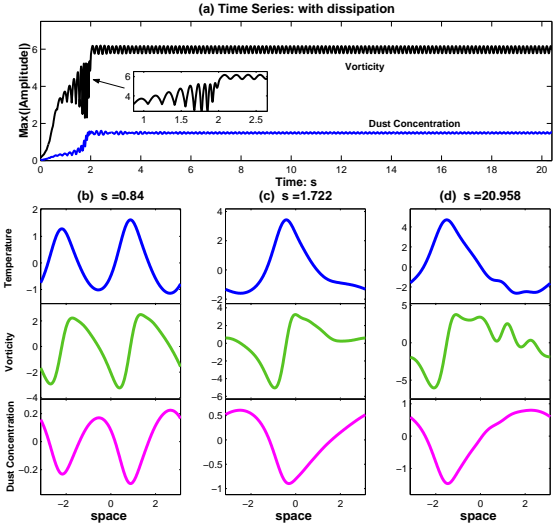}
\end{center}
\caption{{\small Representative profiles for Model C showing
a prograde shear profile:
$r_2 = 5.30$ and $\tilde \phi = 10$, $\tilde\beta = -10$ along with
$\tilde b_1 = -10$.
These results were generated with ${\cal B} = 1$.
(a) Time series of the absolute amplitude of
$F_\chi$ and $\Delta$.  The amplitudes reach a steady average value but
each profile experiences a small amplitude oscillation that neither grows nor
decays with time.  Inset is shown a higher resolution profile of
the transient phase for the vorticity near $s \sim 2$.  The profile
goes from nominally two large scale humps to one.
Panels (b-d) show profiles of various quantities at three
different times:
(b), initial growth phase, time $s\sim 0.84$,
(c), transient adjustment phase, time $s\sim 1.72$,
(d), steady state phase, time $s\sim 20.0$.
In (d) it is clear that the vorticity takes on
significant and complicated structure and it is
also apparent that those
regions where the vorticity is negative are
strongly concentrated.}}
\label{fig:prograde:shear}
\end{figure}

\subsubsection{Model C: Prograde Shear}
\begin{figure}
\begin{center}
\leavevmode
\epsfysize=2.75cm
\epsfbox{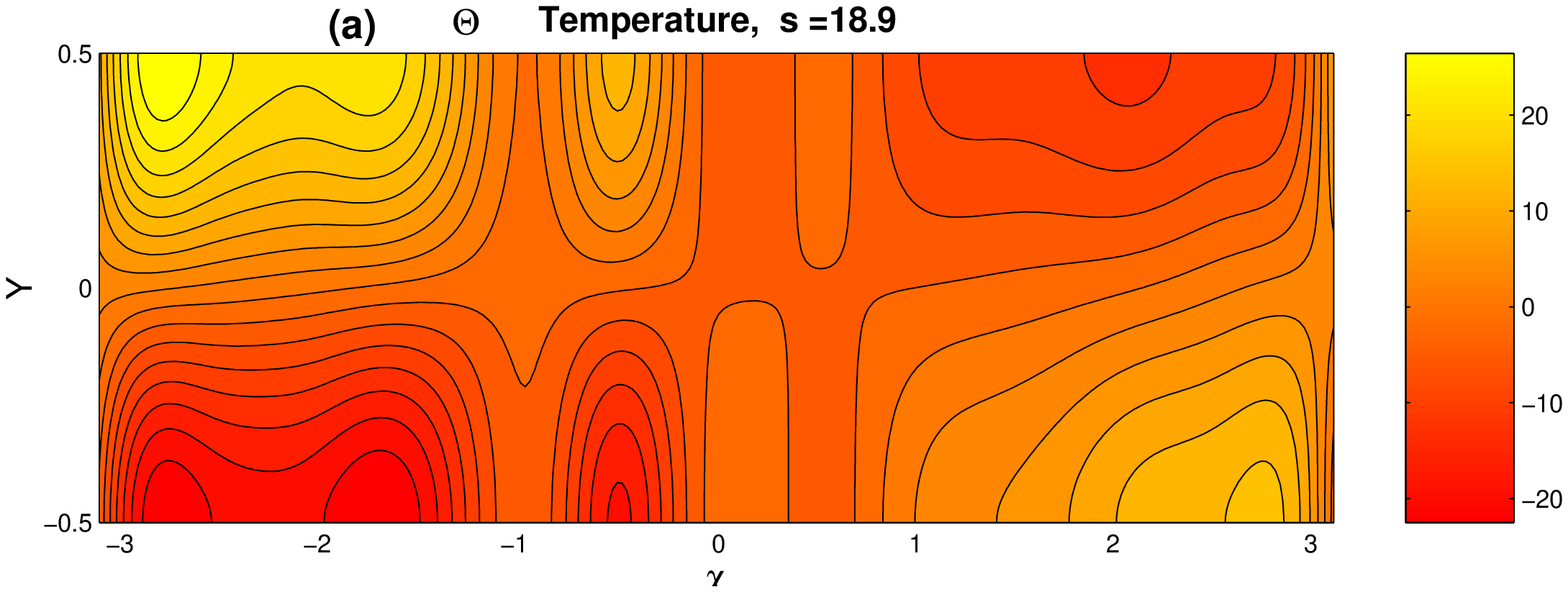}
\end{center}
\label{fig:shear_temperature}
\begin{center}
\leavevmode \epsfysize=2.75cm
\epsfbox{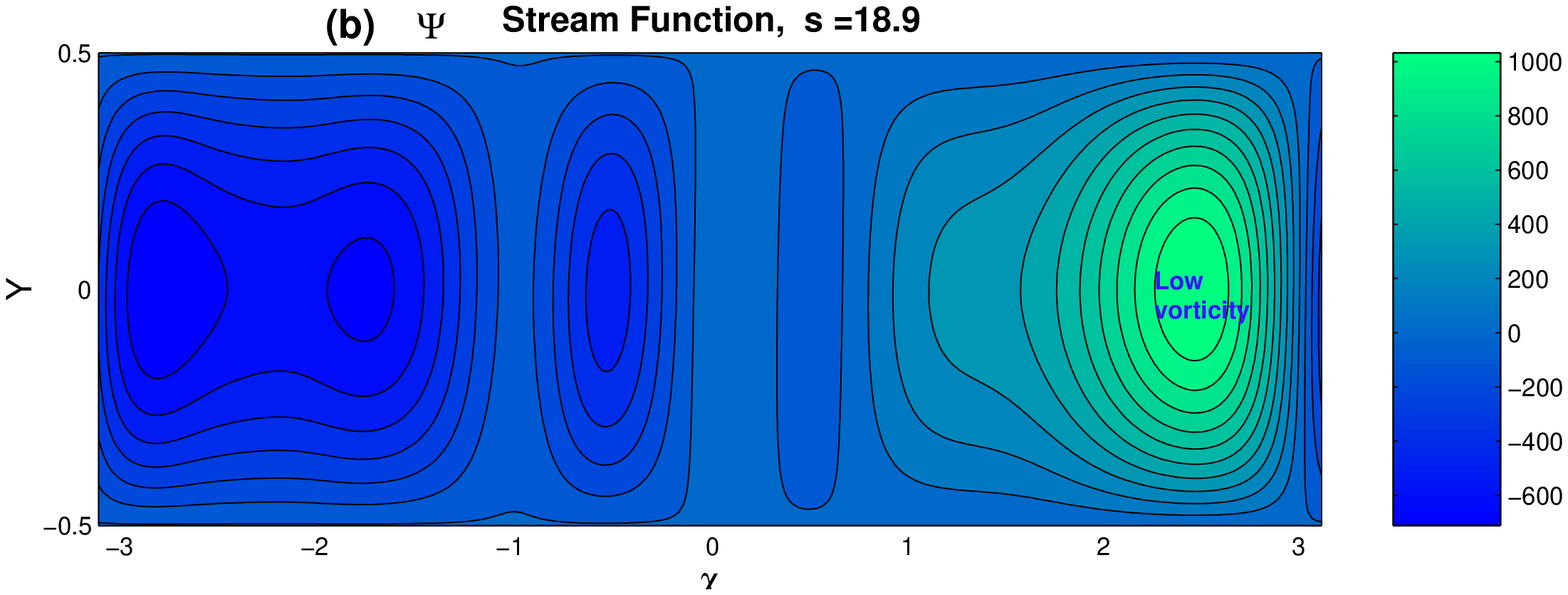}
\end{center}
\caption{{\small Representative contours for Model c
in the quasi-steady state regime, $s \sim 18.9$.
For these contour calculations, $\epsilon = 1$.
(a) Temperature fluctuation contours, $\Theta$.
(b) Stream function contours, $\Psi$.}}
\label{fig:prograde_shear_stream}
\end{figure}

Shear does not change the basic features of the results found
for Model B.  However, it does alter the appearence of the finally
developed convection roll by giving it a much more complicated
structure.  As before, the evolution goes through an initial growth
stage followed by a transient readjustment phase before settling
onto a roughly steady profile (Figure \ref{fig:prograde:shear}a).
The great difference between these results and that of Model B is
that the negative vorticity of the roll is strongly concentrated
into a small region in space whereas the positive vorticity, including
all the substructures like in
Figure \ref{fig:prograde:shear}d, are spread out over larger areas
of space (Figure \ref{fig:prograde_shear_stream}b).

\bigskip
\section*{Acknowledgments}
I would like to thank the DPW at BRC (2002) for providing and
maintaining the desert field facilities in which the
ideas for this work first originated.

\appendix

\section{Asymptotic Expansions}\label{asymptotix}
At infinite Prandtl numbers, the full governing equations
(\ref{ap:mom:body}-\ref{ap:particle:body}) are reproduced here
in slightly different form and with a little bit more detail,
\beqa
R_0\left(1 + \epsilon^2\frac{R_2}{R_0}\right)\partial_X\Theta
- (1+\epsilon^2\delta)\Omega & = & \epsilon^2\beta\partial_X\Psi
\label{ap:mom}
\\
\epsilon^4\partial_T\Theta + \epsilon^2{\cal J}(\Theta,\Psi)
+ \epsilon b_1 z \partial_X\Theta & = &
\epsilon^2\partial_X\Psi + \epsilon^2\partial_X^2\Theta + D^2\Theta
\label{ap:heat}
\\
\Omega &=& \left(D^2 + \epsilon^2\partial_X^2\right)\Psi
\label{ap:stream}
\\
\epsilon^6 C_6 &=& \epsilon^6\phi\Omega + \cdots
\label{ap:particle_divergence}
\\
\Omega_d &=& \epsilon^{11}\frac{1}{\phi}\Omega \\
\epsilon^6\partial_T\delta + \epsilon^6(1+\epsilon^2\delta)C_6
& = & - \epsilon^3 u^{(x)}_d\partial_X\delta
- \epsilon^2 u^{(y)}_dD\delta
\label{ap:particle}
\eeqa
The following expansions are presumed
\beqa
\Theta = \Theta_0 + \epsilon^2\Theta_2 + \epsilon^4\Theta_4 + \cdots \\
\Psi = \Psi_0 + \epsilon^2\Psi_2 + \cdots \\
\Omega = \Omega_0 + \epsilon^2\Omega_2 + \cdots \\
\delta = \delta_2 +  \cdots
\eeqa
and similarly for the other quantities.
At the lowest order (\ref{ap:heat}) is,
\beq
D^2 \Theta_0 = 0
\eeq
which admits the solution $\Theta_0 = f(X,T)$.  Next
solve the next order momentum equation (\ref{ap:mom})
and stream function relationship (\ref{ap:stream})
\beq
R_0\partial_X\Theta_0 - \Omega_0 = 0,
\qquad
\Omega_0 = D^2\Psi_0,
\label{omega:0:sol}
\eeq
together these equations yield the solution
\beq
\Psi_0 = R_0 P_0 f_X, \qquad P_0 = \frac{y^2}{2} - \frac{1}{8},
\label{psi:0:sol}
\eeq
where $P_0$ is designed so that there is no normal flow of
fluid at the vertical boundaries, or in other words,
$\Psi_0 = 0$ at $y = \pm 1/2$.  We note that because of
the relationships between the lowest order vorticity, stream function and
the temperature in (\ref{omega:0:sol}) and
(\ref{psi:0:sol}), the vorticity is proportional to
the gradient of the temperature structure function, or  in other
words, $\Omega_0 \sim F_X$.\par

In order for a solution to exist for the temperature at the next order,
a solvability condition must be
enforced upon the equations
at this order.  This means that starting with,
\beq
D^2\Theta_2 = -\partial_X\Psi_0 - \partial_X^2\Theta_0
+ b_1 y\partial_X\Theta_0
+ \partial_X(\Psi_{0})D\Theta_0 -
 D(\Psi_{0})\partial_X\Theta_0,
\label{ap:heat:2}
\eeq
the solvability condition, in which the thermal flux at
the vertical boundaries is fixed, amounts to requiring,
\beq
\int_{-1/2}^{1/2}D^2\Theta_2dy =
D\Theta_2|^{y=1/2}_{y=-1/2} = 0.
\label{ap:heat:2:solvability}
\eeq
Since $\Theta_0$ is constant with respect to $y$ and since $\Psi_0$ is
an even function of $y$, we find that the condition
(\ref{ap:heat:2:solvability}) applied to
(\ref{ap:heat:2}) results in a choice for $R_0$,
\beq
1+R_0\int_{-1/2}^{1/2}P_0dy = 0.
\eeq
From (\ref{psi:0:sol}), it follows that $R_0 = 12$.
The solution of $\Theta_2$ becomes
\beq
\Theta_2 = T_2f_{XX} + M_2(f_X)^2 + b_1 N_2 f_X,
\eeq
where the individual functions, $T_2,M_2,N_2$ are given by
\beqa
D^2T_2 = -1 - R_0 P_0, &\qquad & T_2 = \frac{y^2}{4}-\frac{y^4}{2}, \\
D^2M_2 =  - R_0D P_0, &\qquad & M_2 = \frac{3y}{2}-2y^3, \\
D^2N_2 =  z, &\qquad & N_2 = -\frac{y}{8}+\frac{y^3}{6}.
\eeqa
these are also designed so
their y-derivatives are zero at the boundaries $y=\pm 1/2$.
Because $u^{(x)}_d$ and $u^{(y)}_d$ are order $\epsilon^7$ and
$\epsilon^6$ respectively,
and furthermore, because of the relationship between
the particle divergence and the gas vorticity in
(\ref{ap:particle_divergence}),
the lowest order expression of (\ref{ap:particle}) is
\beq
\partial_T\delta_2 + \phi\Omega_0 = 0.
\eeq
But because of the solutions (\ref{omega:0:sol}) and
(\ref{psi:0:sol}) we may write
\beq
\delta_2 = \delta_2(X,T),
\eeq
in other words, the structure function $\delta$ for
the particle concentration is independent of $y$ at this
order.  The governing evolution equation for
$\delta_2$ is rewritten to reflect the solution $\Omega_0$,
\beq
\partial_T\delta_2 = -\phi R_0 f_X.
\label{delta:evolution}
\eeq\par
We may now proceed to order $\epsilon^2$ and solve for the
next order stream function $\Psi_2$.  The governing equation
becomes,
\beq
D^2\Psi_2 =
R_0\partial_X\Theta_2 + R_2\partial_X\Theta_0
-\partial_X^2\Psi_0 - \delta_2\Omega_0 - \beta
\partial_X\Psi_0
\eeq
Utilizing the solutions from the previous orders we find
that the $\Psi_2$ may be written as
\beqa
\Psi_2 &=& S_2 f_{XXX} + (R_2 - \delta_2 R_0) P_0 f_X
+ Q_2 \partial_X(f_X)^2 \nonumber \\
& & + \left(b_1 L^{({\rm odd})}_2
- \beta L^{({\rm even})}_2 \right)f_{XX}
\eeqa
where the structure functions appearing above are given
by
\beqa
D^2 S_2 = R_0 T_2 - R_0P_0, &\ &
S_2 = -\frac{y^6}{5}
-\frac{y^4}{4}
+\frac{3y^2}{4}
-\frac{27}{160}, \\
D^2 Q_2 = R_0M_2, &\ &
Q_2 =
-\frac{6y^5}{5}
+3y^3
-\frac{27y}{40} , \\
D^2 L^{({\rm odd})}_2   = R_0N_2,
 &\ &
L^{({\rm odd})}_2 =
\frac{y^5}{10}
-\frac{y^3}{4}
+\frac{9y}{160} , \\
D^2 L^{({\rm even})}_2   = R_0 P_0,
 &\ &
L^{({\rm even})}_2 =
\frac{y^4}{2}
-\frac{y^2}{4}
+\frac{5}{32}.
\eeqa
The structure functions are designed so that
they are zero at the boundaries $y = \pm 1/2$.
Next we turn to the order $\epsilon^4$
heat equation which is,
\beqa
& & \partial_T\Theta_0 + {\cal J}(\Psi_2,\Theta_0)
+{\cal J}(\Psi_0,\Theta_2)
+ b_1 y \partial_X\Theta_2 = \nonumber \\
& & \ \ \ \ \partial_X\Psi_2 + \partial_X^2\Theta_2 + D^2\Theta_4.
\eeqa
Like at the previous order
the solvability condition is,
\beq
\int_{-1/2}^{1/2}D^2\Theta_4dy =
D\Theta_4|^{y=1/2}_{y=-1/2} = 0.
\label{ap:heat:4:solvability}
\eeq
The solvability condition produces an evolution equation
for $f$, namely,
\beqa
& & f_T + \kappa f_{XXXX} - \mu(f^3_X)_X \nonumber \\
& &+ \left[(r_2 - \delta_2)f_{X}\right]_X
+b_1 S (f_X^2)_X + \beta B f_{XXX}
= 0,
\label{f:evolution}
\eeqa
where the constants and parameters are defined by,
\beqa
\kappa &=& -\int^{1/2}_{-1/2}(S_2 + T_2) dy
 =  \frac{2}{21}, \\
\mu &=& -R_0\int^{1/2}_{-1/2}(P_0 DM_2) dy
 = \frac{6}{5}, \\
r_2 &=& -R_2\int^{1/2}_{-1/2}P dy
+b_1^2\int^{1/2}_{-1/2}zN_2 dy
=
\frac{R_2}{R_0} - \frac{b_1^2}{120}, \\
S &=& R_0\int^{1/2}_{-1/2}P_0 DN_2 dy = \frac{1}{10},\\
B &=& R_0\int^{1/2}_{-1/2}L^{({\rm even})}_2 dy = \frac{1}{10}.
\eeqa

\section{Numerical Method}\label{numerical_method}
The general form of the equations we seek to solve are,
\beqa
\frac{\partial F}{\partial s}  &=& {\cal L}F + {\cal N}(F,\Delta) \\
\frac{\partial \Delta}{\partial s}  &=& {\cal M}(F),
\eeqa
where ${\cal L}$ is some linear operator and where
${\cal N}$
is a nonlinear operator involving some combination of
the arguments and
${\cal M}$ is linear in $F$.
\par
These one dimensional equations will be solved in terms of Fourier expansions
of $F$ and $\Delta$ as in,
\beqa
F &=& \sum F_k(s)\exp{ik\chi} + {\rm c.c.} \\
\Delta &=& \sum \Delta_k(s)\exp{ik\chi} + {\rm c.c.},
\eeqa
such that each Fourier component evolves according to
\beqa
\frac{\partial F_k}{\partial s}  &=& {\cal L}_kF_k + {\cal N}_k(F,\Delta) \\
\frac{\partial \Delta_k}{\partial s}  &=& {\cal M}_k(F).
\eeqa
The scalars ${\cal L}_k$ are the linear operator ${\cal L} $
acting on the appropriate
Fourier wave $k$ and where
${\cal N}_k$  and ${\cal M}_k$ are the $k$th components of each respective
operator ${\cal N}$  and ${\cal M}$.
\par
The symbol $F^{n}$ represents the nth time iterate of $F$.  This same
convention carries over to $\Delta$, ${\cal M}$ and ${\cal N}$  .
The evolution is carried out using a second order time centered difference
scheme together with a Crank-Nicholson scheme for the spatial wavenumbers of
the linear operator terms.  All
${\cal N}_k$  and ${\cal M}_k$
expressions are evaluated at the centered time point.
With $\delta s$ as the time increment between time step $n$ and
$n+1$ the discretized equations are,
\beqa
\frac{F^{n+1}_k - F^{n-1}_k}{2 \delta s} &=&
\frac{{\cal L}_k F^{n+1}_k + {\cal L}_k F^{n-1}_k}{2} + {\cal N}^n_k, \\
\frac{\Delta^{n+1}_k - \Delta^{n-1}_k}{\delta s} &=& {\cal M}^n_k .
\eeqa
Rearranging the terms and isolating the $n+1$ time step reveals,
\beqa
F^{n+1}_k &=&
\frac{1+\delta s {\cal L}_k}{1-\delta s {\cal L}_k} F^{n-1}_k
+ \frac{2 \delta s}{1-\delta s {\cal L}_k} {\cal N}^{n}_k ,
\label{disc:p1}
\\
\Delta^{n+1}_k &=& \Delta^{n-1}_k + 2\delta s {\cal M}^n_k.
\label{disc:2}
\eeqa
In the first discretization above, the coefficient terms in front
of $F^{n-1}_k $ and ${\cal N}^n_k$ look like P\'ade approximants
to exponentials of ${\cal L}_k$.  This is then assumed
and the following replacements,
\beq
\frac{1+\delta s {\cal L}_k}{1-\delta s {\cal L}_k}
\rightarrow \exp\left[2\delta s {\cal L}_k\right],
\qquad
\frac{1}{1-\delta s {\cal L}_k}
\rightarrow \exp\left[\delta s {\cal L}_k\right],
\eeq
are made in (\ref{disc:p1}).   The resulting discretized equations that
are simultaneously solved at each time step are (\ref{disc:2}) along with,
\beq
F^{n+1}_k =
\exp\left[2\delta s {\cal L}_k\right] F^{n-1}_k
+ 2 \delta s
\exp\left[\delta s {\cal L}_k\right]
{\cal N}^{n}_k.
\label{disc:1}
\eeq
Time centered differenced schemes like the simple one used here are known
to sometimes incur a period $2\delta s$ long-time numerical instability.
In order to wash this effect out, at every 50th timestep
(\ref{disc:1}) and
(\ref{disc:2}) are advanced through for one time step
using an Euler type evolver as in the following,
\beqa
F^{n+1}_k &=&
\exp\left[\delta s {\cal L}_k\right] F^{n}_k
+ \delta s
\exp\left[\delta s {\cal L}_k\right]
{\cal N}^{n}_k,
\label{euler:disc:1} \\
\Delta^{n+1}_k &=& \Delta^{n}_k + \delta s {\cal M}^n_k.
\label{euler:disc:2}
\eeqa
This Euler advancer is also used as the very first time step of
all simulations.
\end{document}